\documentclass[twocolumn,revtex4,apj,iop, twocolappendix, numberedappendix]{openjournal}
\usepackage{xcolor}
\usepackage{graphicx}
\usepackage{amsfonts}
\usepackage{amssymb, bm}
\usepackage{url}
\usepackage{amsmath}
\usepackage[varg]{txfonts}
\usepackage[normalem]{ulem}
\usepackage{fontawesome}
\usepackage[breaklinks,colorlinks,citecolor=blue,linkcolor=blue,urlcolor=blue]{hyperref}

\DeclareMathAlphabet{\mathbbold}{U}{bbold}{m}{n}
\NewDocumentCommand{\hth}{}{\hat{\theta}}
\NewDocumentCommand{\hsis}{}{|\hat{\sigma}|^2}

\renewcommand{\d}{\mathrm{d}}

\definecolor{MyB}{rgb}{0.1,0.1,1.0}

\setcitestyle{numbers,square}
\begin{document}

\title{Cosmography for a General Spacetime Centred at Arbitrary Redshift}

\author{Jonas Broe Bendtsen$^{\;a,}$\footnote{jonasbb@student.matnat.uio.no}}
\author{Asta Heinesen$^{\;b,c}$\footnote{asta.heinesen@nbi.ku.dk}}
\author{Sofie Marie Koksbang$^{\;a}$ \vspace{0.15cm}\footnote{koksbang@cp3.sdu.dk}}

\affiliation{${}^a$ CP3-Origins, University of Southern Denmark, Campusvej 55, DK-5230 Odense M, Denmark}
\affiliation{${}^b$ Department of Physics \& Astronomy, Queen Mary University of London,  Mile End Road, London E1 4NS, UK\vspace{0.05cm}} 
  \affiliation{${}^c$Niels Bohr Institute, Blegdamsvej 17, DK-2100 
Copenhagen, Denmark}

\begin{abstract}
With upcoming surveys providing large volumes of highly precise observational data across a wide range of redshifts, it is increasingly important to have tools that can translate observational data into geometric and dynamical information without imposing a predetermined cosmological model. General cosmographic expansions centred at arbitrary redshift provide exactly such a tool. We here present the formalism for general cosmographic expansions centred at an arbitrary redshift, valid for 4-dimensional Lorentzian spacetimes. We then apply the expansion formalism to test its ability to reproduce the redshift-distance relation in two examples of large-scale cosmic structures (an underdensity and an overdensity) modelled by the Lema\^itre-Tolman-Bondi metric, where we examine the effect of choosing different redshift intervals for the cosmographic series expansions. This quantifies the extent to which cosmographic coefficients inferred from redshift-distance observations retain their interpretation as local geometric and dynamical quantities, as is expected in standard FLRW cosmology. Similarly to earlier results, we here find that in more general spacetimes the coefficients instead become effective parameters reflecting the finite observational range probed. Lastly, we discuss possible strategies for using the expansions to constrain dynamical and geometric quantities. 
\end{abstract}

\maketitle 
\tableofcontents

\section{Introduction}
Mapping the large-scale structure of the Universe traditionally relies on assuming a specific spacetime, in practice almost exclusively the Friedmann--Lema\^itre--Robertson--Walker (FLRW) spacetime and its perturbative extension, usually restricting attention to the (perturbed) $\Lambda$CDM in particular.
\newline\indent
The most well-known cosmic map is that of the Cosmic Microwave Background (CMB), which can be considered a map of the early-universe density distribution, but there is a long list of other methods used to ``map the Universe''. Some examples are: Weak lensing observations which can be used to map the three dimensional distribution of dark+baryonic matter \cite{weaklensing1}, galaxy-redshift surveys that trace the three dimensional positions of galaxies \cite{galaxy} and maps of the peculiar velocity field that can be obtained through redshift-space distortion or by combining independent measurements of the redshift of, and distance to, astronomical sources \cite{velocity, CosmicFlows}. In case of the latter, a map of the total matter density field can also be inferred. In all these cases, a crucial element in  making the maps is FLRW cosmology and cosmological perturbation theory.
\newline\indent
The cosmological tensions (see e.g. \cite{tension1, tension2, tension3, tension4, tension5} for examples), combined with the rapid improvement in observational data currently being realized, motivate the development of frameworks that minimize model dependence and thus allow data itself to reveal the geometry, dynamics and matter distribution of the Universe. In this work, we present the formalism for general cosmographic expansions centred at arbitrary redshift, and propose to use it to map the Universe with only the minimal assumptions of spacetime being a 4-dimensional Lorentzian manifold, 
and observations fulfilling the notions that (i) light follows null-geodesics and (ii) redshift as a function of affine parameter is invertible.
\newline\indent
Cosmographic expansions are Taylor expansions of observable relations such as the redshift-distance relation, and are standard tools in cosmology routinely used for e.g. inferring $H_0$ \cite{Riess:2016jrr, Riess:2021jrx, Dhawan:2022yws, Galbany:2022zir, Uddin:2023iob, Riess:2022oxy, DES:2024ywx}. The expansions have almost exclusively been considered as expansions centred on $z = 0$, but recently, \cite{arbitraryz1, arbitraryz2} extended the standard framework to include expansions centred on arbitrary redshift. Here, it was demonstrated how expanding around different redshifts can be used to obtain tight constraints on $H(z)$. Yet, the formalism remains model-dependent and applicable only in the strictly homogeneous and isotropic regime of observations because the expansions are based on the FLRW metric assumption.
\newline\indent
Parallel to the development and use of these standard FLRW-based cosmographic expansions, there has also been an advancement of constructing \cite{umeh2013, Clarkson:2011uk, Heinesen:2020bej, Maartens:2023tib, Kalbouneh:2024szq} and employing \cite{Dhawan:2022lze, Cowell:2022ehf, Macpherson:2025qec} cosmographic expansions that apply to more general inhomogeneous spacetimes, known as {\em inhomogeneous}, {\em general} or {\em covariant} cosmography. We extend these efforts by presenting a general cosmography centred around an arbitrary redshift. Our motivation for this extension is twofold. First, several studies \cite{convergence1, convergence2, convergence3} demonstrate that general cosmographic expansions may exhibit a limited radius of convergence (relative to the Hubble scale) and/or slow convergence, causing low-order truncations to deviate significantly from the exact redshift-distance relation in spacetimes containing structure. This has important implications for the physical interpretation of the expansion coefficients of the cosmography and the scale they probe \cite{Macpherson:2025qec,convergence2}. By allowing the cosmographic formalism to be centred at different redshifts, one may mitigate these convergence issues by stitching together multiple local expansions, each valid in its own redshift interval. Second, centring the expansion at other redshifts, enables the fitted coefficients to directly probe the local properties around the chosen expansion point.
\newline\newline
After presenting the cosmographic expansion in section \ref{sec: theory}, we demonstrate its use by applying it to two Lema\^itre--Tolman--Bondi \cite{Lemaitre1933,Lemaitre1997,Tolman1934,Bondi1947} (LTB) models. The models are defined in section \ref{sec:LTBtheory} and results of applying the cosmographic framework to redshift-distance relations in the models are presented in section \ref{sec:LTBresults} where we comment on the accuracy of the expansions and how this affects the interpretation of cosmographic expansion coefficients. We discuss our results in section \ref{sec:discussion} and summarize and conclude in section \ref{sec:conclusion}. We set $c = 1$ throughout.

\section{General Cosmography centred at arbitrary redshift}\label{sec: theory}
For an arbitrary spacetime, we may consider the Taylor expansion of the angular diameter distance about an arbitrary redshift, $z_*$, and write
\begin{align}
\begin{split}
d_A(z)&\approx d_A(z_*) + d'_A(z_*)(z-z_*)     \\&+ \frac{1}{2} d_A''(z_*)(z-z_*)^2 + \frac{1}{6}d_A'''(z_*)(z-z_*)^3,
\end{split}
\end{align}
where a prime denotes derivatives with respect to the redshift. We now wish to obtain the expressions for $d_A', d_A'', d_A'''$ in terms of geometric and dynamical spacetime quantities. For this, it will be useful to remember the relations \cite{god_bog}
\begin{align}
    \frac{\d z}{\d \lambda} &= - (1+z)^2 E_o\mathcal{H}\\
    \frac{\d \hth}{\d \lambda} &= -\frac{1}{2} \hth^2 - 2\hsis - k^\mu k^\nu {R_{\mu\nu}} \\
    \frac{\d d_A}{\d\lambda} &= \frac{1}{2} \hth d_A,
\end{align}
where $E = -u^\mu k_\mu$ is the photon energy and evaluation at the observer is indicated by a subscripted $o$. We have chosen the convention $\hat\theta = \nabla_\mu k^\mu$, where $\hat\theta$ is the isotropic expansion of the image. $R_{\mu\nu}$ is the Ricci curvature, $\hsis = \frac{1}{2}\hat{\sigma}^{\mu\nu} \hat{\sigma}_{\mu\nu}$ is the shear, representing anisotropic image distortion, and $\lambda$ is the affine parameter along a light ray with tangent vector $k^\mu$. We have defined the line-of-sight expansion rate
\begin{align}
    \mathcal{H} = \frac{1}{3}\theta -e^\mu a_\mu + e^\mu e^\nu \sigma_{\mu\nu},
\end{align}
where $\theta$ is the local expansion rate of the fluid, $a^\mu$ is its 4-acceleration, $\sigma_{\mu\nu}$ its shear tensor and $e^\mu$ represents the spatial direction of observation of an observer comoving with the fluid.
\newline\indent
Using the above, we readily see that
\begin{equation}
    d_A' = - \frac{1}{(1+z)^2 E_o\mathcal{H}} \frac{\d d_A}{\d \lambda} = - \frac{\hat\theta}{2(1+z)^2 E_o\mathcal{H}} d_A
    \label{eq:1st}
\end{equation}
and
\begin{equation}
    d_A'' = -\bigg[\frac{2}{(1+z)} + \frac{\mathcal{H}'}{\mathcal{H}}\bigg] d_A' - \frac{2 \hsis + k^\mu k^\nu R_{\mu\nu}}{2(1+z)^4E_o^2\mathcal{H}^2} d_A .
    \label{eq:2nd_diffs}
\end{equation}
The latter can be expanded as
\begin{equation}
    d_A'' = \frac{d_A}{2(1+z)^2 E_o\mathcal{H}}\bigg[ \frac{2\hat\theta}{(1+z)} + \frac{\hat\theta\mathcal{H}'}{\mathcal{H}}  - \frac{2 \hsis + k^\mu k^\nu R_{\mu\nu}}{(1+z)^2 E_o\mathcal{H}} \bigg] ,
    \label{eq:2nd}
\end{equation}
in agreement with the results of \cite{PRL}. We now extend the analysis of \cite{PRL} and introduce the third-order derivative of $d_A$, finding
\begin{multline}
    d_A'''(z) = - \bigg[\frac{2}{1+z} + \frac{\mathcal{H}'}{\mathcal{H}}\bigg] d_A'' \\- \bigg[\frac{2 \hsis + k^\mu k^\nu R_{\mu\nu}}{2(1+z)^4 E_o^2 \mathcal{H}^2} - \frac{2}{(1+z)^2} + \frac{\mathcal{H}''}{\mathcal{H}} - \frac{\mathcal{H}'^2}{\mathcal{H}^2} \bigg] d_A' \\ - \frac{1}{(1+z)^4 E_o^2 \mathcal{H}^2} \Bigg[ \frac{2 \hth \hsis + k^\alpha k^\beta C_{\mu\alpha\nu\beta} \hat{\sigma}^{\mu\nu}}{(1+z)^2 E_o \mathcal{H}} +  \frac{\big(k^\mu k^\nu R_{\mu\nu}\big)'}{2} \\ - \big(2 \hsis + k^\mu k^\nu R_{\mu\nu}\big) \bigg(\frac{2}{1+z} + \frac{\mathcal{H}'}{\mathcal{H}} \bigg) \Bigg] d_A ,
    \label{eq:3rd_diffs}
\end{multline}
or, with the derivatives of $d_A$ written out explicitly,
\begin{multline}
d_A'''(z) = \frac{d_A}{2(1+z)^2 E_o \mathcal{H}} \Bigg[ \frac{3 \big(2 \hsis + k^\mu k^\nu R_{\mu\nu}\big)}{(1+z)^2 E_o \mathcal{H}} \bigg[ \frac{2}{1+z} + \frac{\mathcal{H}'}{\mathcal{H}} \bigg]\\
+ \frac{\hat{\theta}\mathcal{H}''}{\mathcal{H}} - \frac{2\hat{\theta}\mathcal{H}'^2}{\mathcal{H}^2} - \frac{4\hat{\theta}\mathcal{H}'}{(1+z)\mathcal{H}} -\frac{6\hat{\theta}}{(1+z)^2} - \frac{\big(k^\mu k^\nu R_{\mu\nu}\big)'}{(1+z)^2 E_o \mathcal{H}} \\
+ \frac{\hat{\theta}\big(2 \hsis + k^\mu k^\nu R_{\mu\nu}\big) - 4\big(2 \hat{\theta} \hsis + k^\alpha k^\beta C_{\mu\alpha\nu\beta} \hat{\sigma}^{\mu\nu}\big)}{2(1+z)^4 E_o^2 \mathcal{H}^2}  \Bigg] .
\label{eq:3rd}
\end{multline}
This expression is formidable, with a variety of terms related to shear, Weyl curvature, line-of-sight expansion, light beam expansion and Ricci focusing. However, in the limit where shear and Weyl curvature are small (expected to be true on large scales in the real universe) and general relativity is assumed to hold, the expression reduces dramatically, and the remaining terms collapse to combinations depending only on the isotropic expansion rate of the beam and line-of-sight, and the matter content, together with their derivatives along the geodesics.
\newline\indent
The FLRW limit of the expressions for $d_A', d_A''$ and $d_A'''$ is shown in appendix~\ref{app:FLRWlimit}.

\subsection{Mapping between expansions}
It is straightforward to convert the Taylor series expansion of $d_A$ into Padé approximants \cite{pade} as done in, e.g., \cite{Adamek:2024hme} or to expansions of the ``expansion rate fluctuation field'' introduced in \cite{Kalbouneh:2022tfw}. We may also use the Taylor expansion of $d_A$ to derive an expansion of the luminosity distance, $d_L$, assuming that the Etherington reciprocity relation \cite{Etherington:1933} holds.
\newline\indent
As a demonstration of mapping from an expansion of $d_A$ to the mapping of other quantities, we will consider mapping between $d_A$ and $d_L$. For this, we first consider the relations
\begin{align}
    \begin{split}
        d_A(z)& = \sum_{n=0}^{\infty}\frac{a_n}{n!}\Delta z^n\\
        (1+z)^2 &= (1+z_*)^2 + 2(1+z_*)\Delta z + \Delta z^2 = c_0 + c_1\Delta z + c_2\Delta z^2\\
        d_L(z)& = (1+z)^2\sum_{n=0}^{\infty}\frac{a_n}{n!}\Delta z^n = \sum_{n=0}^{\infty}\frac{b_n}{n!}\Delta z^n\\&
        = \left((1+z_*)^2 + 2(1+z_*)\Delta z + \Delta z^2\right)\sum_{n=0}^{\infty}\frac{a_n}{n!}\Delta z^n
,    \end{split}
\end{align}
where we have introduced $\Delta z:= z-z_*$ and $c_0, c_1$ and $c_2$ are used as shorthand notation for the Taylor expansion coefficients of $(1+z)^2$. Comparing the first and last expressions for $d_L$ we see that 
\begin{align}
    \begin{split}
        b_0 &= c_0a_0 = (1+z_*)^2a_0\\
        b_1 &= c_0a_1 + c_1a_0 = (1+z_*)^2a_1 + 2(1+z_*)a_0\\
        b_2&= c_0a_2 + 2c_1a_1 + 2c_2a_0 = (1+z_*)^2a_2 + 4(1+z_*)a_1 + 2a_0\\
        b_3 & = c_0a_3 + 3c_1a_2 + 6c_2a_1 = (1+z_*)^2a_3 + 6(1+z_*)a_2 + 6 a_1,
    \end{split}
\end{align}
where $a_0 = d_A(z_*), a_1 = d_A'(z_*), a_2 = d_A''(z_*)$ and $a_3 = d_A'''(z_*)$. We thus obtain the expansion
\begin{align}
\begin{split}
    d_L(z)&\approx (1+z_*)^2d_A(z_*) + \left( (1+z_*)^2 d_A'(z_*) + 2(1+z_*)d_A(z_*) \right)\Delta z \\&+ \frac{1}{2}\left( (1+z_*)^2d_A''(z_*) + 4(1+z_*)d_A'(z_*) + 2d_A(z_*) \right)\Delta z^2\\ & + \frac{1}{6}\left((1+z_*)^2d_A'''(z_*)+ 6(1+z_*)d_A''(z_*) + 6d_A'(z_*)\right)\Delta z^3,
\end{split}
\end{align}
valid sufficiently close to $z = z_*$. For an explicit and detailed demonstration of transformation from a Taylor expansion to a Padé approximant, the reader is referred to \cite{arbitraryz2}.

\section{General Cosmography in LTB Models: Theoretical Background}\label{sec:LTBtheory}
\begin{figure}
    \centering
    \includegraphics[width=0.7\linewidth]{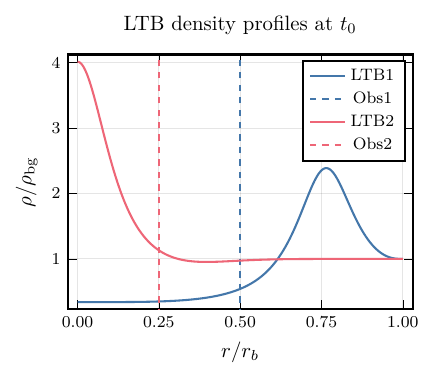}
    \caption{Present-time density profiles of the two LTB models as a function of the radius $r$ scaled to the boundary radius $r_b$. Dashed, vertical lines indicate positions of observers used in section \ref{sec:LTBresults}.}
    \label{fig:rho_r}
\end{figure}

We will demonstrate the use of our new cosmographic expansion by comparing its approximations with the exact redshift-distance relation in Lema\^itre--Tolman--Bondi (LTB) models  \cite{Lemaitre1933,Lemaitre1997,Tolman1934,Bondi1947}. The LTB model is a spherically symmetric inhomogeneous dust+$\Lambda$ spacetime. It typically consists of a spherical void surrounded by a compensating overdense region such that the surrounding spacetime reduces exactly or asymptotically to an FLRW ``background'', but can also be used to model a central overdensity. The line element of the LTB model can be written as
\begin{equation}
    \d s^2 = -\d t^2 + \frac{A_{,r}(t, r)^2}{1-k(r)} \d r^2 + A(t, r)^2 \d \Omega^2,
\end{equation}
where $k(r)$ is a curvature function and $A(t,r)$ is the areal radius.
\newline\indent
We will consider two concrete LTB models. Both models have a vanishing big bang time, meaning that coordinate time is also cosmic time. The first model, LTB1, represents a central void surrounded by a mass-compensating overdensity, and is further specified by introducing the curvature function as
\begin{equation}
    k(r) = \begin{cases}
        -r^2 k_\mathrm{max}\big((r/r_b)^n - 1\big)^m & r\leq r_b \\
        0 & r > r_b
    \end{cases},
\end{equation}
where $r_b$ is the boundary of the LTB region, outside which the model reduced exactly to a $\Lambda$CDM model. The constant $k_\mathrm{max}$ defines the magnitude of the curvature, while $m$ and $n$ are even positive integers. To specify LTB1, we set $m = 4 = n$, $k_{\rm max} = 5.4\times 10^{-8}$ and $r_b = 40$Mpc.
\newline\indent
The second model, LTB2, represents a central overdensity and has $k(r)$ given by
\begin{align}
    k(r) = k_0r^2\delta(r),
\end{align}
where 
\begin{align}
    \delta(r) = \delta_0\exp(-r^2/2/r_\sigma^2).
\end{align}
We use $k_0 = \frac{5}{3} (a_i H_i)^2$,  $\delta_0 = 10^{-3}$, $r_\sigma = 7$Mpc. In our notation, $a_i, H_i$ are the scale factor and Hubble rate of the background FLRW spacetime at some initial time, as detailed below.
\newline\indent
The function $A(t,r)$ is obtained by integrating the $rr$ component of Einstein's field equations for the LTB spacetime, which gives the following relation \cite{Bolejko2010}
\begin{equation}\label{eq:Adot}
    \frac{dA^2(t,r)}{dt} = -k(r) + \frac{2M(r)}{A(t,r)} + \frac{\Lambda}{3}A(t,r)^2,
\end{equation}
where $M(r)$ is a function of integration and the dot denotes temporal derivative. $M(r)$ is interpreted as the enclosed effective gravitational mass inside a shell of radius $r$. For an FLRW universe, $M(r)$ is given by
\begin{equation}\label{eq:M_FLRW}
    M_\mathrm{FLRW} = \frac{4\pi G_N}{3}a(t)^3 r^3 \rho_\mathrm{FLRW}(t).
\end{equation}
$M(r)$ is constant in time since $\rho \propto a^{-3}$. Physically, this is because mass does not disappear from the comoving shells. In the LTB case with vanishing cosmological constant we can approximate $M(r)$ as \cite{VanAcoleyen:2008cy}
\begin{equation}
    M(r) = \frac{4\pi G_N}{3}a_i^3 r^3 \rho_{\mathrm{FLRW},i} \Bigg(1 + \frac{3}{5}\frac{k(r)}{\big(a_i H_i r\big)^2}\Bigg),
\end{equation}
where the subscript $i$ is used to denote an initial value. We choose to set initial values at cosmic time $t(a = 1/1200)$ in the background FLRW model, i.e. in the FLRW model which the LTB model reduces to in its homogeneous region. We choose our background model to be the flat $\Lambda$CDM model with $\Omega_{m,0} = 0.30$ and $H_0 = 70.0\,\mathrm{km}\,\mathrm{s}^{-1}\,\mathrm{Mpc}^{-1}$. Since initial conditions are set at early times, we can neglect $\Lambda$ when setting initial conditions and use the expression for $M(r)$ introduced above. We now have both $k(r)$ and $M(r)$ analytically and can find $A(t,r)$ at any time once we have set its initial conditions, using $A(t_i,r) = ra_i$.
\newline\newline
Figure \ref{fig:rho_r} shows the present-time density profile of the two density distributions.

\subsection{Light propagation in LTB models}
To obtain exact redshift-distance relations in the LTB models, we solve the geodesic equations with initial conditions fulfilling the null condition, $k^\mu k_\mu = 0$. Simultaneously with solving the geodesic equations, we solve the transport equation for the tidal matrix to obtain the deformation matrix/Jacobian $D$ with $d_A = \sqrt{
|\det(D)|}$. See e.g. \cite{Koksbang:2021zyi} for details. We set initial conditions using $k^t = -1$ and spatial components $k^i$ determined from a \texttt{HEALpix}\footnote{https://healpix.sourceforge.io/} \cite{healpix} skymap distribution as follows: We generate a set of angular sky coordinates $(\theta_{\rm HP}, \phi_{\rm HP})$ using \texttt{HEALpix}. We the project these to local coordinates on the observer's sky using
\begin{align}
\begin{split}
    \hat p_r &= \sin\theta_0\sin\theta_{\rm HP}\cos(\phi_0 - \phi_{\rm HP}) + \cos\theta_0\cos\theta_{\rm HP}\\
    \hat p_\theta & = \cos\theta_0\sin\theta_{\rm HP}\cos(\phi_0- \phi_{\rm HP}) - \sin\theta_0\cos\theta_{\rm HP}\\
    \hat p_\phi & = -\sin\theta_{\rm HP}\sin(\phi_0-\phi_{\rm HP}),
\end{split}
\end{align}
where $\phi_0, \theta_0$ are the angular coordinates of the observer in the local LTB coordinate system. This relation was obtained by taking the dot products between the \texttt{HEALpix} unit vectors and the observer's unit position vector.
We then calculate the corresponding $k^i$ by using the null condition to find initial conditions
\begin{align}
    k^i= k^t \hat p_i \sqrt{-g_{tt}/g_{ii}}.
\end{align}
Since the Weyl distortion has earlier been found by one of the authors to be highly sub-dominant to the Ricci contribution in this type of models, we omit the former here for simplicity. This means that the tidal matrix can be computed simply using the Ricci tensor and that we do not need to set initial conditions for and parallel transport the screen space basis vectors.

\section{General Cosmography in LTB Models: Numerical examples}\label{sec:LTBresults}
\begin{figure}
    \centering
    \includegraphics[width=0.7\linewidth]{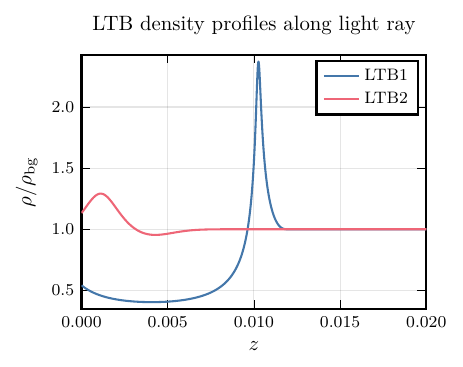}
    \caption{The density of the two LTB models along a fiducial light ray.}
    \label{fig:rho_z}
\end{figure}

\begin{figure}
    \centering
    \includegraphics[width=0.8\linewidth]{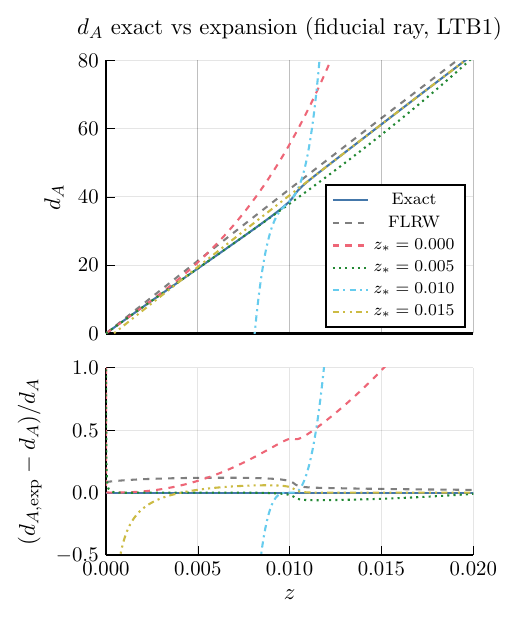}
    \caption{Cosmographic expansions of $d_A$ in LTB1 centred at
    $z_*=0.000$, $z_*=0.005$, $z_*=0.010$, and $z_*=0.0015$ compared to the exact solution of a fiducial light ray.}
    \label{fig:expansion_LTB1}
\end{figure}

\begin{figure}
    \centering
    \includegraphics[width=0.8\linewidth]{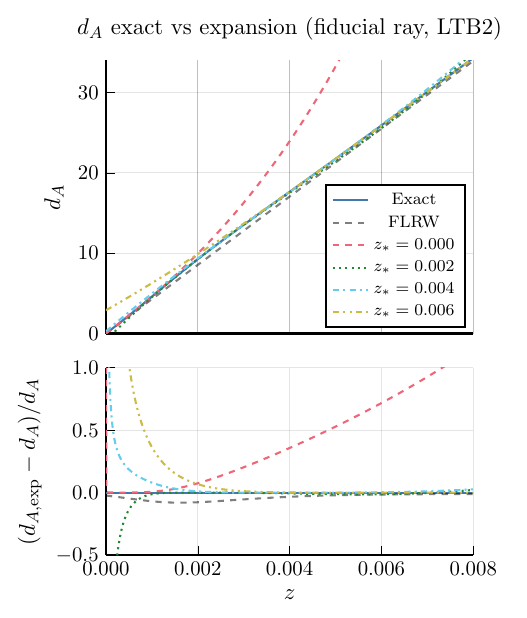}
    \caption{Cosmographic expansions of $d_A$ in LTB2 centred at
    $z_*=0.000$, $z_*=0.002$, $z_*=0.004$, and $z_*=0.0006$ compared to the exact solution of a fiducial light ray.}
    \label{fig:expansion_LTB2}
\end{figure}

We consider one observer in each LTB model. Each observer is placed in a region with large gradient of the density field in order to maximize the effects of the structures. The positions of the observers are indicated by vertical lines in Figure~\ref{fig:rho_r}.
\newline\newline
To qualitatively asses the accuracy of the Taylor expansion up to third order, we first consider the angular diameter distance along a fiducial light ray in each model. The densities along each fiducial ray are shown in figure \ref{fig:rho_z}. The resulting cosmographic expansions are shown in figures \ref{fig:expansion_LTB1} and \ref{fig:expansion_LTB2} where they are compared with the exact angular diameter distance along the fiducial rays. In these figures, we include cosmographic expansions centred on four different redshifts. For LTB1 we centre the expansions on $z_* = 0, 0.005, 0.010, 0.015$ while we use $z_* = 0, 0.002, 0.004, 0.006$ for LTB2. Figures \ref{fig:expansion_LTB1} and \ref{fig:expansion_LTB2} clearly demonstrate that the accuracy of the Taylor expansion drops dramatically when moving only slightly away from $z_*$. The approximation is poorest when the expansion is centred in a region with steep density gradient such as $z_*  = 0.01$ in LTB1.


\begin{figure*}
    \centering

    \begin{minipage}{0.24\textwidth}
        \centering
        \includegraphics[width=\linewidth]{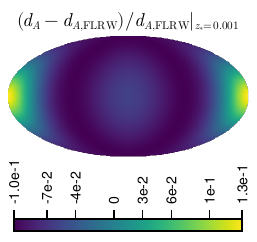}
    \end{minipage}
    \hfill
    \begin{minipage}{0.24\textwidth}
        \centering
        \includegraphics[width=\linewidth]{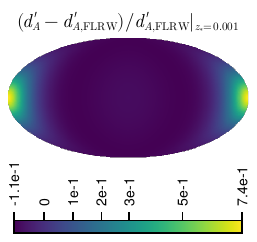}
    \end{minipage}
    \hfill
    \begin{minipage}{0.24\textwidth}
        \centering
        \includegraphics[width=\linewidth]{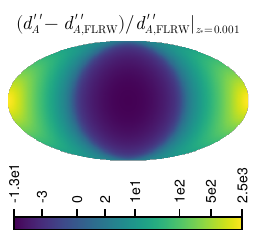}
    \end{minipage}
    \hfill
    \begin{minipage}{0.24\textwidth}
        \centering
        \includegraphics[width=\linewidth]{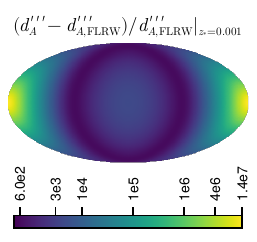}
    \end{minipage}

    \vspace{0.5cm}
    
    \begin{minipage}{0.24\textwidth}
        \centering
        \includegraphics[width=\linewidth]{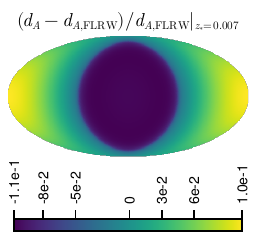}
    \end{minipage}
    \hfill
    \begin{minipage}{0.24\textwidth}
        \centering
        \includegraphics[width=\linewidth]{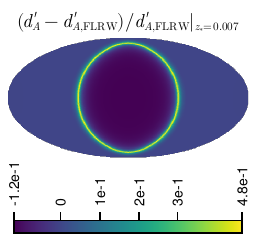}
    \end{minipage}
    \hfill
    \begin{minipage}{0.24\textwidth}
        \centering
        \includegraphics[width=\linewidth]{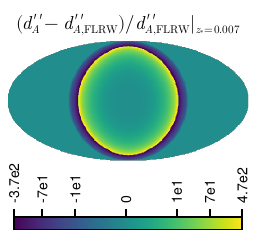}
    \end{minipage}
    \hfill
    \begin{minipage}{0.24\textwidth}
        \centering
        \includegraphics[width=\linewidth]{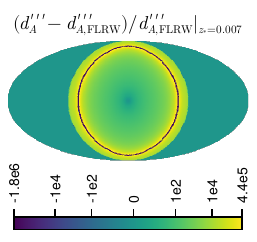}
    \end{minipage}

    \vspace{0.5cm}

    \begin{minipage}{0.24\textwidth}
        \centering
        \includegraphics[width=\linewidth]{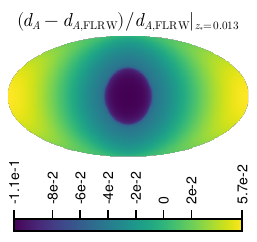}
    \end{minipage}
    \hfill
    \begin{minipage}{0.24\textwidth}
        \centering
        \includegraphics[width=\linewidth]{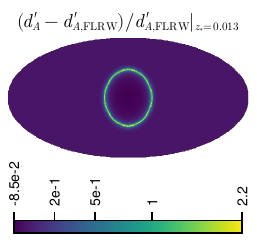}
    \end{minipage}
    \hfill
    \begin{minipage}{0.24\textwidth}
        \centering
        \includegraphics[width=\linewidth]{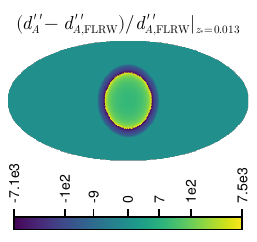}
    \end{minipage}
    \hfill
    \begin{minipage}{0.24\textwidth}
        \centering
        \includegraphics[width=\linewidth]{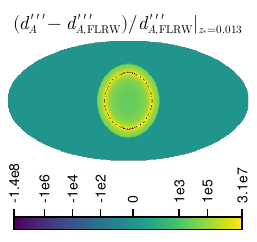}
    \end{minipage}

    \vspace{0.5cm}

    \begin{minipage}{0.24\textwidth}
        \centering
        \includegraphics[width=\linewidth]{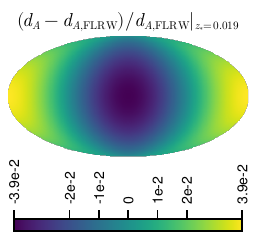}
    \end{minipage}
    \hfill
    \begin{minipage}{0.24\textwidth}
        \centering
        \includegraphics[width=\linewidth]{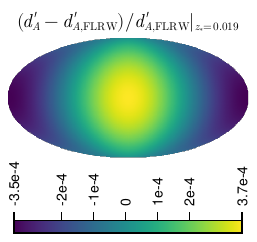}
    \end{minipage}
    \hfill
    \begin{minipage}{0.24\textwidth}
        \centering
        \includegraphics[width=\linewidth]{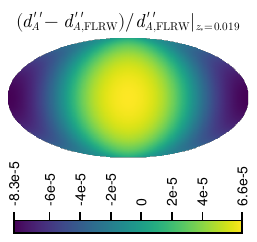}
    \end{minipage}
    \hfill
    \begin{minipage}{0.24\textwidth}
        \centering
        \includegraphics[width=\linewidth]{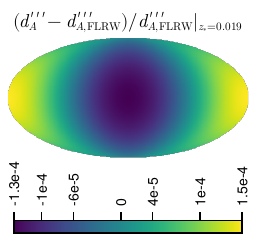}
    \end{minipage}
    \caption{Skymaps of the LTB1 model showing the relative deviation of the angular diameter distance compared to the FLRW limit. From left to right, the plots show $d_A$, $d_A'$, $d_A''$, and $d_A'''$. The skymaps are evaluated around different redshifts, from top to bottom: $z_*= 0.001$, $z_*= 0.007$, $z_*= 0.013$, and $z_*= 0.019$. An asinh scale is used for the colorbar.}
    \label{fig:skymaps1}
\end{figure*}

\begin{figure*}
    \centering

    \begin{minipage}{0.24\textwidth}
        \centering
        \includegraphics[width=\linewidth]{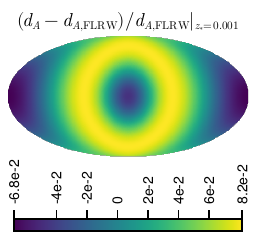}
    \end{minipage}
    \hfill
    \begin{minipage}{0.24\textwidth}
        \centering
        \includegraphics[width=\linewidth]{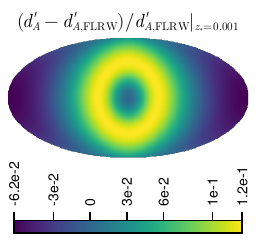}
    \end{minipage}
    \hfill
    \begin{minipage}{0.24\textwidth}
        \centering
        \includegraphics[width=\linewidth]{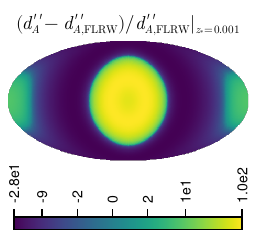}
    \end{minipage}
    \hfill
    \begin{minipage}{0.24\textwidth}
        \centering
        \includegraphics[width=\linewidth]{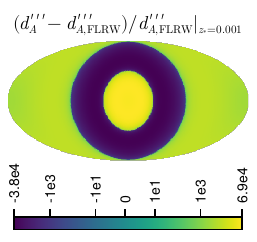}
    \end{minipage}

    \vspace{0.5cm}
    
    \begin{minipage}{0.24\textwidth}
        \centering
        \includegraphics[width=\linewidth]{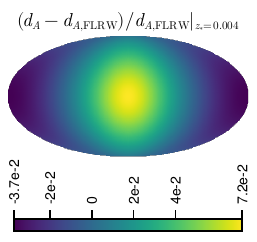}
    \end{minipage}
    \hfill
    \begin{minipage}{0.24\textwidth}
        \centering
        \includegraphics[width=\linewidth]{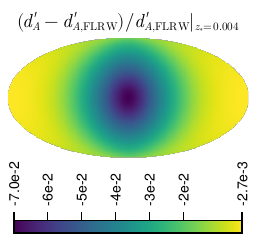}
    \end{minipage}
    \hfill
    \begin{minipage}{0.24\textwidth}
        \centering
        \includegraphics[width=\linewidth]{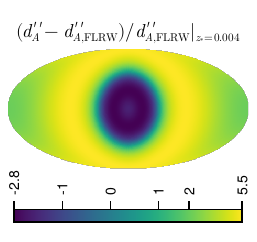}
    \end{minipage}
    \hfill
    \begin{minipage}{0.24\textwidth}
        \centering
        \includegraphics[width=\linewidth]{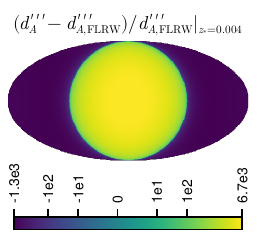}
    \end{minipage}

    \vspace{0.5cm}

    \begin{minipage}{0.24\textwidth}
        \centering
        \includegraphics[width=\linewidth]{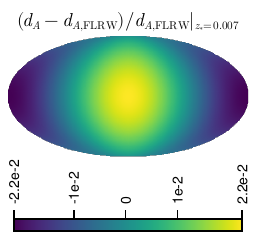}
    \end{minipage}
    \hfill
    \begin{minipage}{0.24\textwidth}
        \centering
        \includegraphics[width=\linewidth]{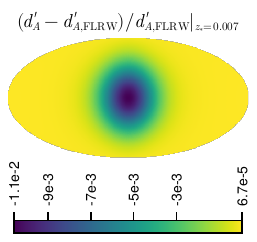}
    \end{minipage}
    \hfill
    \begin{minipage}{0.24\textwidth}
        \centering
        \includegraphics[width=\linewidth]{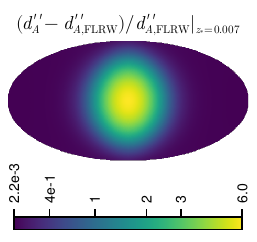}
    \end{minipage}
    \hfill
    \begin{minipage}{0.24\textwidth}
        \centering
        \includegraphics[width=\linewidth]{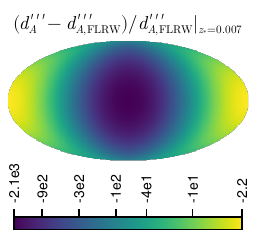}
    \end{minipage}

    \vspace{0.5cm}

    \begin{minipage}{0.24\textwidth}
        \centering
        \includegraphics[width=\linewidth]{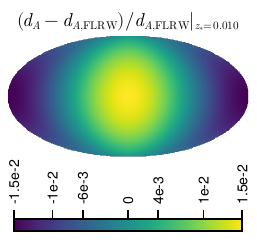}
    \end{minipage}
    \hfill
    \begin{minipage}{0.24\textwidth}
        \centering
        \includegraphics[width=\linewidth]{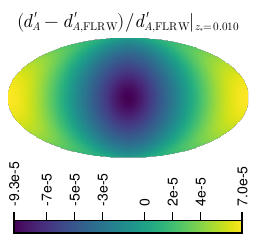}
    \end{minipage}
    \hfill
    \begin{minipage}{0.24\textwidth}
        \centering
        \includegraphics[width=\linewidth]{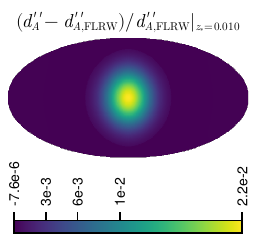}
    \end{minipage}
    \hfill
    \begin{minipage}{0.24\textwidth}
        \centering
        \includegraphics[width=\linewidth]{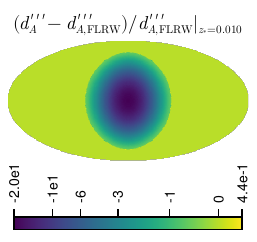}
    \end{minipage}
    \caption{Skymaps of the LTB2 model showing the relative deviation of the angular diameter distance compared to the FLRW limit. From left to right, the plots show $d_A$, $d_A'$, $d_A''$, and $d_A'''$. The skymaps are evaluated around different redshifts, from top to bottom: $z_*= 0.001$, $z_*= 0.004$, $z_*= 0.007$, and $z_*= 0.010$. An asinh scale is used for the colorbar.}
    \label{fig:skymaps2}
\end{figure*}

\subsection{Skymaps of expansion coefficients}
We now move on to consider multiple light rays for the two observers. Figures \ref{fig:skymaps1} and \ref{fig:skymaps2} show skymaps of the expansion coefficients, $d_A(z_*), d_A'(z_*), d_A''(z_*)$ and $d_A'''(z_*)$ for each observer. The maps were made by using healpy\footnote{https://healpy.readthedocs.io/en/latest/} with 49152 light rays and are dispalyed using the Mollweide projection. Figure \ref{fig:skymaps1} shows the expansion coefficients compared to their background values for the observer in the LTB1 model at four different redshifts. One of the main features of these skymaps is the clear distinction between inside and outside the inhomogeneous region. For instance, in the top row ($z_* = 0.001$), the light rays are almost exclusively emitted from the underdense region. The only exception is the yellow areas at the sides of the figures. At the two next redshifts, disk-like shapes with $d_A$ values below the background value appear. These again correspond to light rays emitted inside the void region. At the fourth redshift, the deviations from the background angular diameter distance is again smaller, in agreement with the rays mainly being emitted from the FLRW region. Near the sharp density contrast, the magnitude of the coefficients increases significantly with order, reducing the effectiveness of the $\Delta z<<1$ expansion. This may indicate either slow convergence of the Taylor series or a reduced radius of convergence. We expect that the dominant effect in this regime is slow convergence and hence that additional terms are needed to achieve an accurate approximation of the true redshift-distance relation. Lastly, note that the more extreme values of the coefficients are located in a narrow ring along the density gradients, while the remaining parts of the skymap are fairly similar to the FLRW values.
\newline\indent
Figure \ref{fig:skymaps2} shows the expansion coefficients for the LTB2 observer at four different values of redshift. The skymaps largely have opposite signs compared to the skymaps for LTB1, simply because LTB2 represents an overdensity rather than a central underdensity. More importantly, the skymaps are significantly smoother than those for LTB1 because LTB2 has smoother density gradients. This means that higher-order derivatives do not take on nearly as extreme values as in the LTB1 case, which we expect increases the rate of convergence of the Taylor series.

\begin{figure*}
    \centering
    
    \begin{minipage}{0.48\textwidth}
        \centering
        \includegraphics[width=\linewidth]{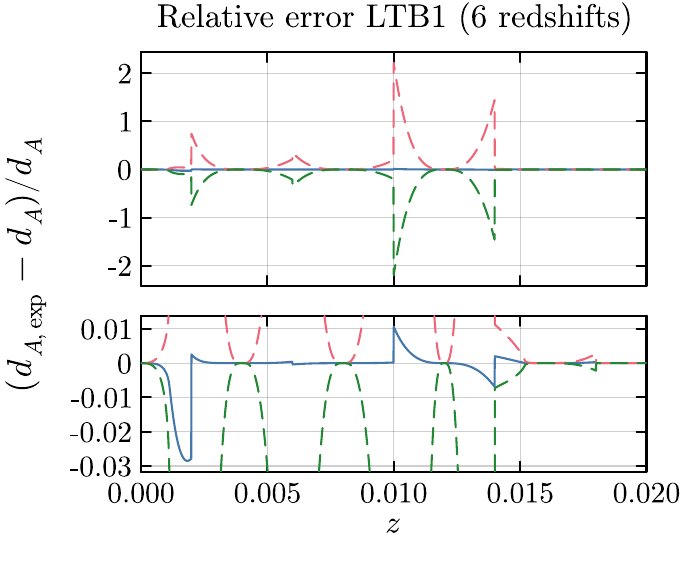}
    \end{minipage}\hfill
    \begin{minipage}{0.48\textwidth}
        \centering
        \includegraphics[width=\linewidth]{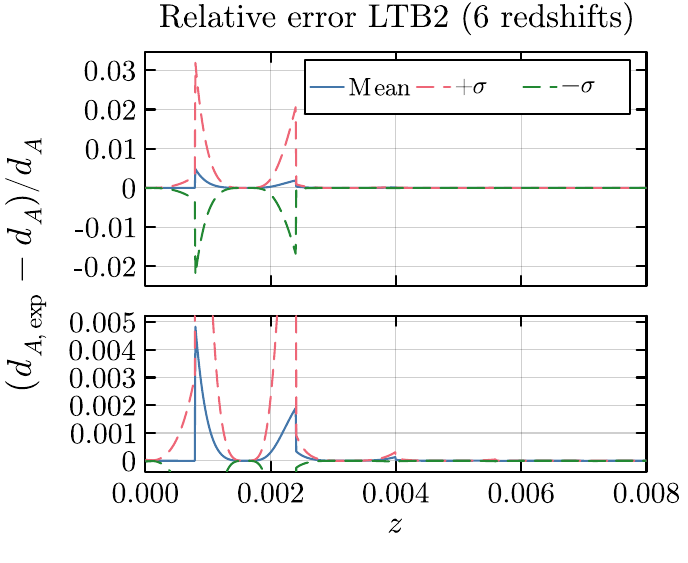}
    \end{minipage}
    
    \vspace{1em}
    
    \begin{minipage}{0.48\textwidth}
        \centering
        \includegraphics[width=\linewidth]{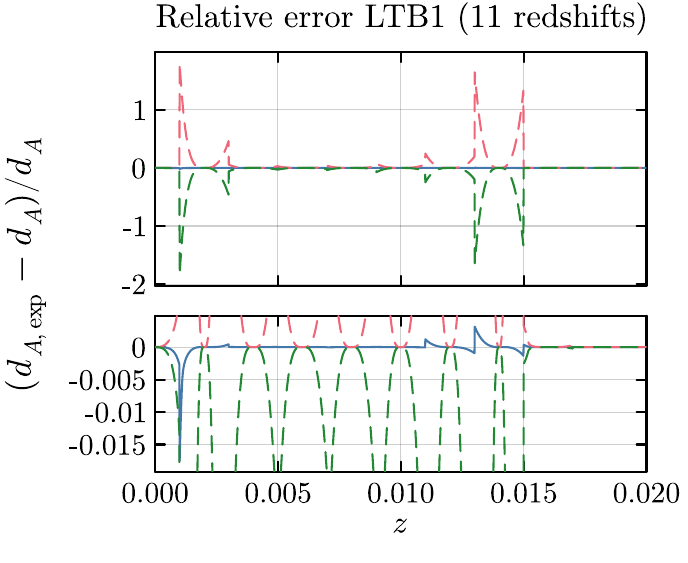}
    \end{minipage}\hfill
    \begin{minipage}{0.48\textwidth}
        \centering
        \includegraphics[width=\linewidth]{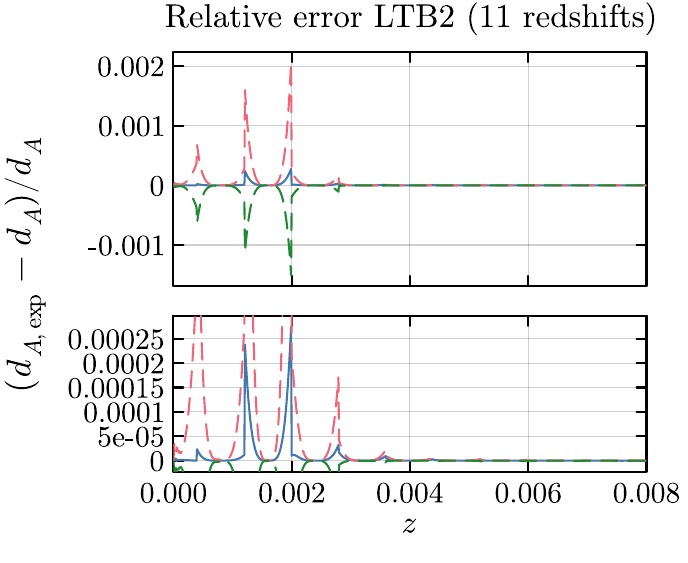}
    \end{minipage}
    
    \vspace{1em}
    
    \begin{minipage}{0.48\textwidth}
        \centering
        \includegraphics[width=\linewidth]{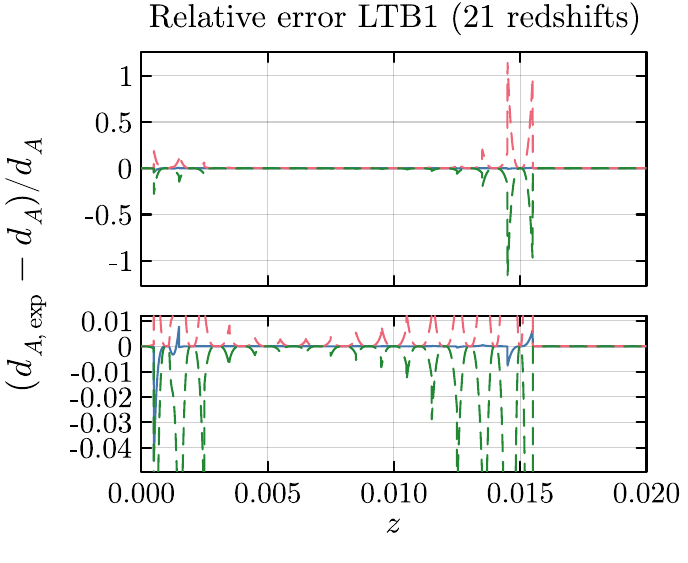}
    \end{minipage}\hfill
    \begin{minipage}{0.48\textwidth}
        \centering
        \includegraphics[width=\linewidth]{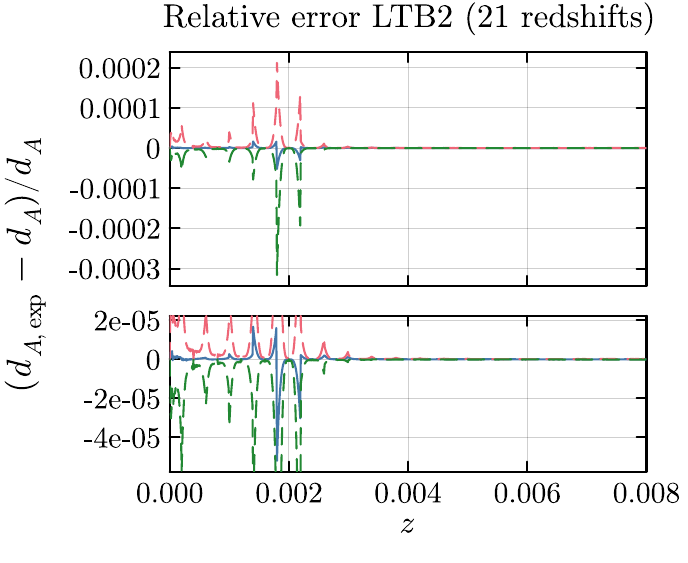}
    \end{minipage}

    \caption{Relative error of the cosmographic expansions for LTB1 (left) and LTB2 (right) for 6, 11 and 21 numbers of expansion points, respectively, from top to bottom. The solid line indicates mean while the dashed lines show the standard deviation $\pm\sigma$ of the mean relative error. Zoom-ins are presented below the main plots to better show the mean relative error.}
    \label{fig:relative_error}
\end{figure*}

\subsection{Piecewise cosmographic expansion}
\begin{figure*}
    \centering

    \begin{minipage}{0.24\textwidth}
        \centering
        \includegraphics[width=\linewidth]{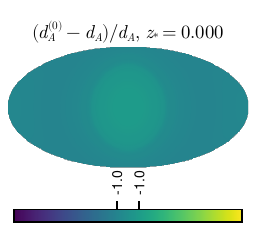}
    \end{minipage}\hfill
    \begin{minipage}{0.24\textwidth}
        \centering
        \includegraphics[width=\linewidth]{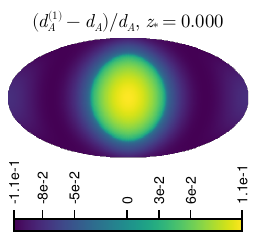}
    \end{minipage}\hfill
    \begin{minipage}{0.24\textwidth}
        \centering
        \includegraphics[width=\linewidth]{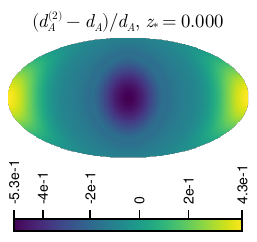}
    \end{minipage}\hfill
    \begin{minipage}{0.24\textwidth}
        \centering
        \includegraphics[width=\linewidth]{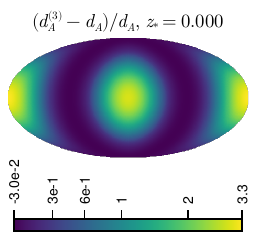}
    \end{minipage}

    \vspace{0.5cm}
    
    \begin{minipage}{0.24\textwidth}
        \centering
        \includegraphics[width=\linewidth]{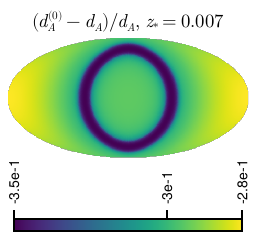}
    \end{minipage}\hfill
    \begin{minipage}{0.24\textwidth}
        \centering
        \includegraphics[width=\linewidth]{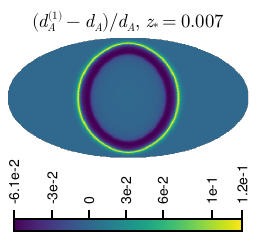}
    \end{minipage}\hfill
    \begin{minipage}{0.24\textwidth}
        \centering
        \includegraphics[width=\linewidth]{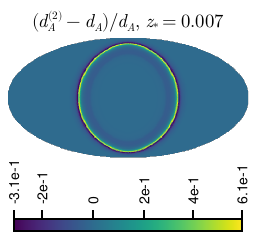}
    \end{minipage}\hfill
    \begin{minipage}{0.24\textwidth}
        \centering
        \includegraphics[width=\linewidth]{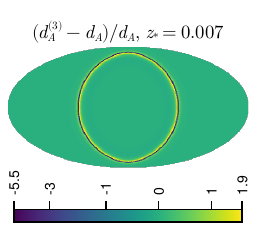}
    \end{minipage}

    \vspace{0.5cm}

    \begin{minipage}{0.24\textwidth}
        \centering
        \includegraphics[width=\linewidth]{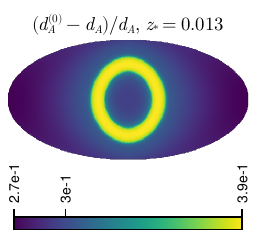}
    \end{minipage}\hfill
    \begin{minipage}{0.24\textwidth}
        \centering
        \includegraphics[width=\linewidth]{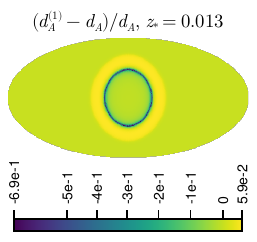}
    \end{minipage}\hfill
    \begin{minipage}{0.24\textwidth}
        \centering
        \includegraphics[width=\linewidth]{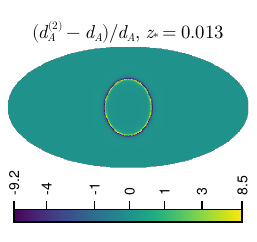}
    \end{minipage}\hfill
    \begin{minipage}{0.24\textwidth}
        \centering
        \includegraphics[width=\linewidth]{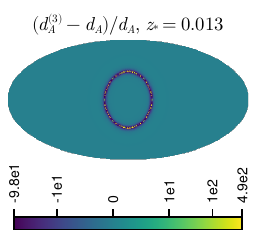}
    \end{minipage}

    \vspace{0.5cm}

    \begin{minipage}{0.24\textwidth}
        \centering
        \includegraphics[width=\linewidth]{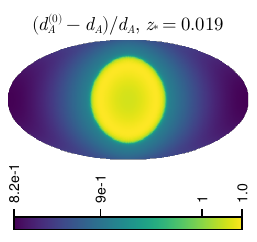}
    \end{minipage}\hfill
    \begin{minipage}{0.24\textwidth}
        \centering
        \includegraphics[width=\linewidth]{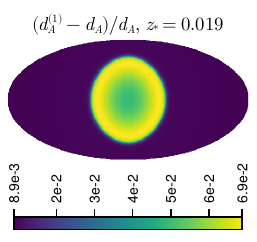}
    \end{minipage}\hfill
    \begin{minipage}{0.24\textwidth}
        \centering
        \includegraphics[width=\linewidth]{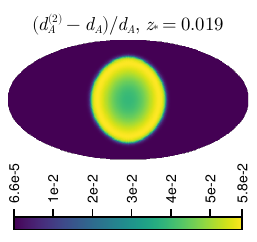}
    \end{minipage}\hfill
    \begin{minipage}{0.24\textwidth}
        \centering
        \includegraphics[width=\linewidth]{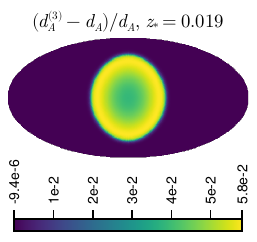}
    \end{minipage}
    
    \caption{Skymaps of LTB1 evaluated at $z=0.010$. The plots show the relative error of the cosmographic expansion compared to the exact $d_A$ to the zeroth, first, second, and third orders (from left to right). The cosmographic expansion is performed around the following redshifts (from top to bottom): $z_*= 0.000$, $z_*= 0.007$, $z_*= 0.013$, and $z_*= 0.019$. Note that $d_A^{(0)}(0) = 0$ so deviations from 1 in the top left plot are numerical artefacts. The colorbar uses an asinh scale.}
    \label{fig:skymaps1_expansion}
\end{figure*}

\begin{figure*}
    \centering

    \begin{minipage}{0.24\textwidth}
        \centering
        \includegraphics[width=\linewidth]{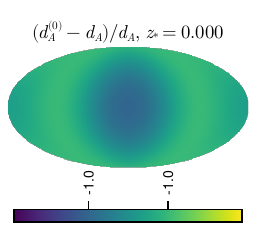}
    \end{minipage}\hfill
    \begin{minipage}{0.24\textwidth}
        \centering
        \includegraphics[width=\linewidth]{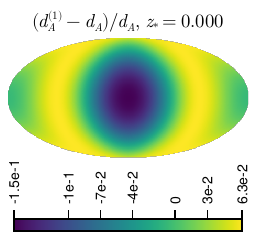}
    \end{minipage}\hfill
    \begin{minipage}{0.24\textwidth}
        \centering
        \includegraphics[width=\linewidth]{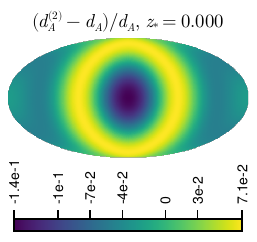}
    \end{minipage}\hfill
    \begin{minipage}{0.24\textwidth}
        \centering
        \includegraphics[width=\linewidth]{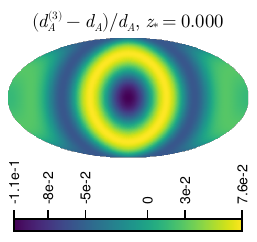}
    \end{minipage}

    \vspace{0.5cm}
    
    \begin{minipage}{0.24\textwidth}
        \centering
        \includegraphics[width=\linewidth]{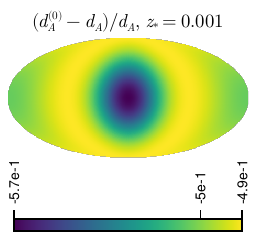}
    \end{minipage}\hfill
    \begin{minipage}{0.24\textwidth}
        \centering
        \includegraphics[width=\linewidth]{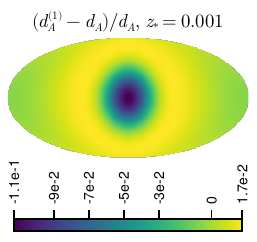}
    \end{minipage}\hfill
    \begin{minipage}{0.24\textwidth}
        \centering
        \includegraphics[width=\linewidth]{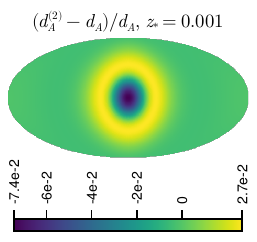}
    \end{minipage}\hfill
    \begin{minipage}{0.24\textwidth}
        \centering
        \includegraphics[width=\linewidth]{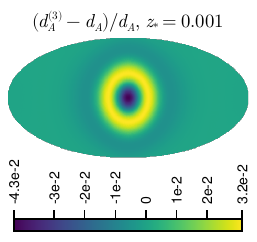}
    \end{minipage}

    \vspace{0.5cm}

    \begin{minipage}{0.24\textwidth}
        \centering
        \includegraphics[width=\linewidth]{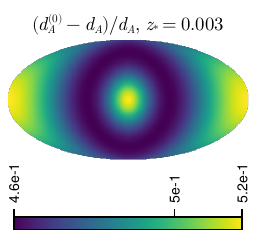}
    \end{minipage}\hfill
    \begin{minipage}{0.24\textwidth}
        \centering
        \includegraphics[width=\linewidth]{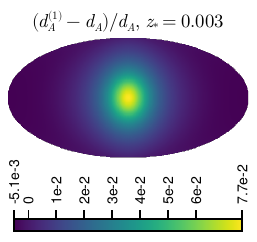}
    \end{minipage}\hfill
    \begin{minipage}{0.24\textwidth}
        \centering
        \includegraphics[width=\linewidth]{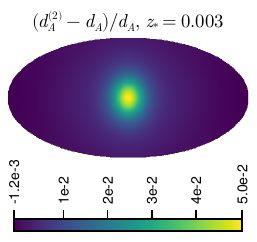}
    \end{minipage}\hfill
    \begin{minipage}{0.24\textwidth}
        \centering
        \includegraphics[width=\linewidth]{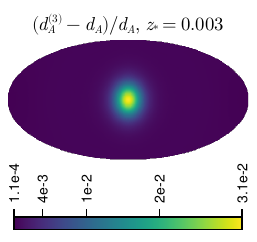}
    \end{minipage}

    \vspace{0.5cm}

    \begin{minipage}{0.24\textwidth}
        \centering
        \includegraphics[width=\linewidth]{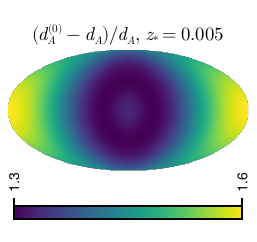}
    \end{minipage}\hfill
    \begin{minipage}{0.24\textwidth}
        \centering
        \includegraphics[width=\linewidth]{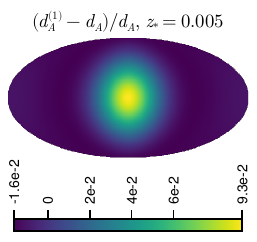}
    \end{minipage}\hfill
    \begin{minipage}{0.24\textwidth}
        \centering
        \includegraphics[width=\linewidth]{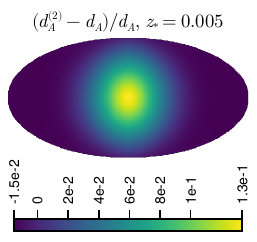}
    \end{minipage}\hfill
    \begin{minipage}{0.24\textwidth}
        \centering
        \includegraphics[width=\linewidth]{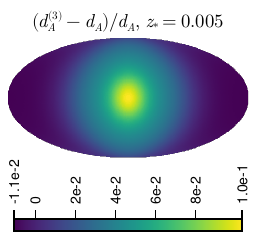}
    \end{minipage}
    
    \caption{Skymaps of LTB2 evaluated at $z=0.002$. The plots show the relative error of the cosmographic expansion compared to the exact $d_A$ to the zeroth, first, second, and third orders (from left to right). The cosmographic expansion is performed around the following redshifts (from top to bottom): $z_*= 0.000$, $z_*= 0.001$, $z_*= 0.003$, and $z_*= 0.005$. Note that $d_A^{(0)}(0) = 0$ so deviations from 1 in the top left plot are numerical artefacts. The colorbar uses an asinh scale}
    \label{fig:skymaps2_expansion}
\end{figure*}

Figures \ref{fig:expansion_LTB1} and \ref{fig:expansion_LTB2} indicate that cosmographic expansions will break down when expansions are performed close to large density gradients. This in in agreement with the skymaps in figures \ref{fig:skymaps1} and \ref{fig:skymaps2} where the expansion coefficients are seen to take very large values in these regions. We will now consider the usefulness of introducing piecewise cosmographic expansions where we stitch together expansions around multiple $z_*$ values using the same 49152 light rays used for making the earlier skymaps. Figure \ref{fig:relative_error} shows the relative errors, $(d_{A,\rm exp}-d_A)/d_A$, where $d_A$ is the exact angular diameter distance obtained through solving the geodesic equations and transport equation, and $d_{A, \rm exp}$ is the corresponding cosmographic approximation at 3rd order.
\newline\indent
The mean relative error is expected to decrease with an increase in expansion points. This is clearly demonstrated in the case of LTB2 as shown in figure \ref{fig:relative_error}, where a doubling of expansion points leads to around a factor ten decrease of the maximum mean relative error. In the case of LTB1, however, the maximum mean relative error does not generally decrease with more expansion points, but stays relatively constant. This is likely due to the higher gradients in this model, seen around $z=0.010$ in figure \ref{fig:expansion_LTB1}. At most redshift values, the higher number of expansion points do, however, improve the mean relative error. The standard deviation spikes are reduced slightly in LTB1 and significantly in LTB2 with increasing number of expansion points, as shown in figure \ref{fig:relative_error}.
\newline\indent
To further explore the piecewise cosmographic expansion we plot a series of skymaps at fixed redshift ($z = 0.01$ for LTB1 and $z = 0.002$ for LTB2), but where the cosmographic expansion is based on different expansion centres. The results are shown in figures  \ref{fig:skymaps1_expansion} and \ref{fig:skymaps2_expansion}. Considering first Figure \ref{fig:skymaps1_expansion} for LTB1, we note that when using $z_* = 0.019$, we see an improvement in the cosmographic reconstruction of the exact angular diameter distance when we go to higher expansion order. Note that for $z_* = 0.019$, the expansion coefficients are evaluated in the FLRW region. However, the skymaps are shown at $z = 0.01$ which for a large portion of the light rays corresponds to inside the inhomogeneous region. For the other considered $z_*$, the improvement (at $z = 0.01$) when going to higher order in the cosmographic expansion is less persistent. For instance, looking at $z_* = 0.013$ we see that the error decreases when going from zeroth to first order. But when going to second and third order, the error increases again. A similar trend is seen for $z_* = 0.007$ and $z_* = 0$. We interpret this as meaning that we are in these cases evaluating the expansion coefficients in regions where the gradients are sufficiently large to affect the convergence properties of the expansion. As a result, the radius of convergence may have been reduced below the point where we compare the expansion and exact redshift-distance relation, or the convergence may have become significantly slower, indicating that more expansion terms are required before the expansion approaches the exact redshift-distance relation in a stable manner. Nonetheless, it is worth noting that the relative difference between the exact redshift-distance relation and the cosmographic expansions are mostly no larger than of order 10 percent, and the larger errors appear to be concentrated in clearly defined circles in the skymaps, corresponding to particularly strong density contrasts.
\newline\indent
In figure \ref{fig:skymaps2_expansion} for the LTB2 model we see a similar trend of not achieving consistent, stable convergence of the cosmographic expansions towards the exact redshift-distance relations. However, in this case, the deviation between the exact and approximate relations are significantly less and the relative deviations are nowhere larger than around 10 \% in the studied skymaps. 
\newline\newline
In summary, our results show that the convergence properties of the general cosmographic expansions are primarily dictated by the local behaviour of the underlying density field and the corresponding geometry of spacetime. Thus, the accuracy of third-order cosmographic approximations is poorest when the expansion is centred at redshifts associated with strong density gradients, while substantially higher accuracy is obtained for similar redshift ranges when the density profile is more smooth. Consequently, piecewise cosmographic reconstructions of the angular diameter distance are not automatically advantageous. Their effectiveness depends critically on the choice of expansion centres, with the greatest benefit obtained when each expansion is centred in a region with sufficiently well-behaved gradients. This highlights the inherent limitation that regions with strong density gradients remain the most difficult to approximate accurately, regardless of the placement of the expansion centres.

\section{Discussion}\label{sec:discussion} 
\begin{figure*}
    \centering
    \includegraphics[width=0.9\linewidth]{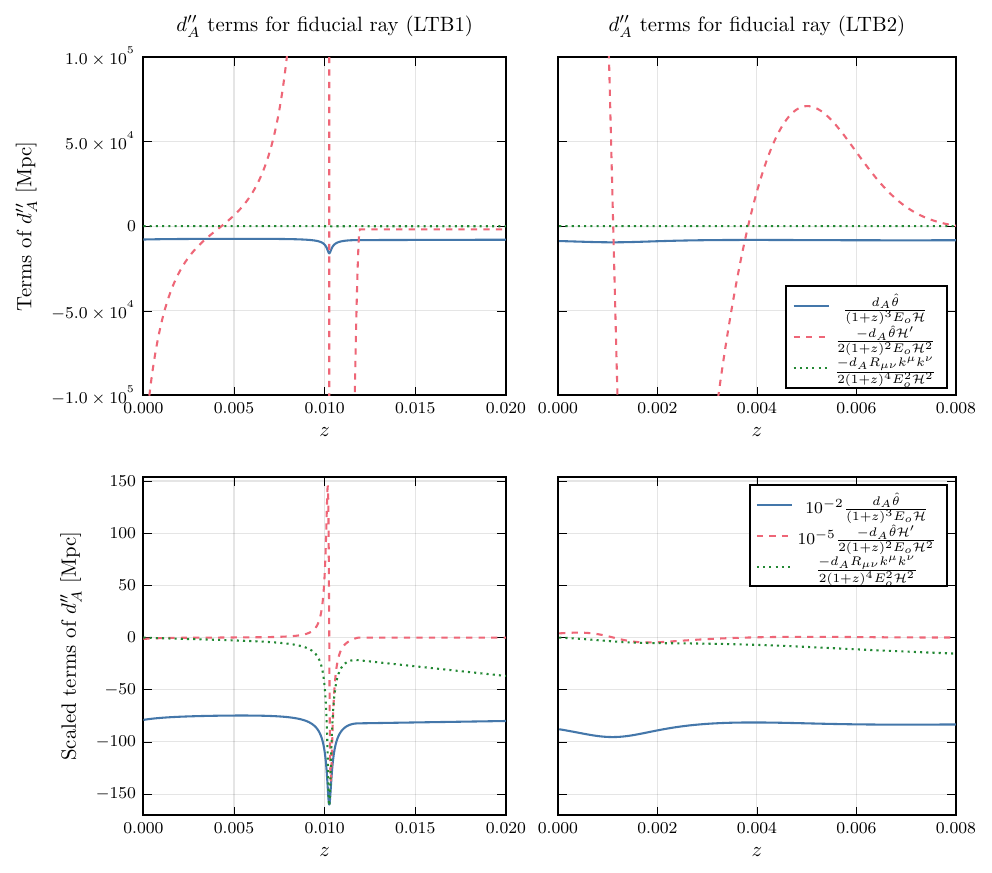}
    \caption{The non-vanishing terms in the second order derivative of $d_A$ shown for the two LTB models. In the bottom row, the $\frac{d_A \hat{\theta}}{(1+z)^3 E_o \mathcal{H}}$ term (blue solid line) is multiplied with $10^{-2}$, while $-\frac{d_A \hat{\theta}\mathcal{H}'}{2(1+z)^2 E_o \mathcal{H}^2}$ is multiplied with $10^{-5}$ (red dashed line) to show all terms clearly in the same plot. The last term $-\frac{d_A R_{\mu\nu}k^\mu k^\nu}{2(1+z)^4 E_o^2 \mathcal{H}^2}$ (green dotted line) is not rescaled, and is thus the least dominant term in these models. The plots share the y-axis and the legend.
    }
    \label{fig:da_zz_terms}
\end{figure*}

In the previous section, we focused on the ability of the cosmographic expansions to approximate the underlying true redshift-distance relation. We did this because the usefulness of cosmographic expansions extends beyond simply being flexible (i.e. polynomial) fits to observational data of the redshift-distance relation. A low-order polynomial may provide an excellent empirical fit of the observed redshift-distance relation over the redshift range sampled by the data. However, this alone does not guarantee that the inferred coefficients correspond to the Taylor coefficients of the underlying redshift-distance relation, since the fitting procedure generally absorbs contributions from neglected higher order terms into the lower order coefficients and since the best fit polynomial may not correspond to the Taylor series of the redshift-distance relation when this is not convergent. In practice, one may assess the robustness of the inferred coefficients by increasing the order of the polynomial fit and verifying that the lower-order coefficients remain stable, indicating that truncation errors are under control. The physical interpretation of the coefficients as measures of the local geometry and dynamics of the Universe at some scale requires the additional assumption that the observed redshift-distance relation is well described by the Taylor expansion of the exact redshift-distance relation in the suitably averaged spacetime at that scale. Here, the physical scale over which the spacetime must be averaged for optimal convergence properties is an important consideration (see e.g. \cite{Macpherson:2025qec,convergence2} for earlier discussions of this point). As a rule of thumb, this scale is set by the mean redshift of the cosmological survey \cite{Macpherson:2025qec}.
\newline\indent
Our study cannot determine the radius of convergence of the expansions, and moreover cannot disentangle a low radius of convergence from slow convergence. However, our study can quantify the accuracy of the third-order Taylor series for the examples studied. This is ultimately the practically relevant quantification for cosmographic data analyses, as it characterizes the accuracy that can be expected from finite-order expansions irrespective of whether their limitations arise from slow convergence or from a small radius of convergence. LTB models provide a controlled setting for investigating the convergence properties of the cosmographic expansions. By considering LTB models with varying degrees of inhomogeneity, we can quantify how departures from homogeneity affect the convergence and hence the order of expansion required for the truncated Taylor expansion to accurately represent the underlying true redshift-distance relation. Such analyses thereby provide guidance on the range of validity we can expect for low-order cosmographic expansions applied to real observational data, and thus whether a given truncation order is sufficient for the fitted coefficients to retain their physical interpretation. Our study of piecewise cosmography provides the main lesson that the convergence behaviour depends less on the partitioning of the redshift range than on where the expansion centres are placed relative to the fluctuations in the density field.
\newline\newline
Another important property of cosmographic expansions is that, when centred at $z = 0$, they admit a finite multipole representation. For example, the third-order expansion of the luminosity distance terminates after a finite number of multipoles, corresponding to only 61 independent degrees of freedom \cite{Heinesen:2020bej}. This finite parameter space makes the expansion tractable and substantially improves the prospects for constraining the coefficients observationally. In the framework presented here, however, where the expansion is centred at an arbitrary redshift, this finite multipole structure is lost. Every multipole can, in principle, contribute to the Taylor expansion, and the spherical harmonic hierarchy generally no longer truncates at any finite order. Consequently, although the multipole formalism is both elegant and efficient for expansions centred at $z = 0$, it becomes considerably less powerful for expansions about an arbitrary redshift.
\newline\indent
The expansions centred around arbitrary redshift constructed here are nevertheless interesting. First of all, the expansion allows the development of controlled, local approximations to the redshift-distance relation {\em anywhere} along a given geodesic, not just at the observer. This makes it possible to probe the influence of local inhomogeneities and anisotropies robustly at any redshift, rather than relying on extrapolations around $z = 0$. Second, by avoiding a multipole decomposition entirely, the method is automatically agnostic to the angular structure of spacetime up to smoothing necessary due to only having a finite amount of data: anisotropies and inhomogeneities are incorporated implicitly through the expansion coefficients themselves.
Finally, constructing these redshift-local expansions opens the door to new observational applications such as piecewise distance reconstruction, consistency checks at specific redshifts, and FLRW model-independent cross-checks against cosmological probes that are naturally centred at $z>0$, such as BAO measurements.
\newline\indent
Overall, expansions of observables around non-zero redshifts offer multiple intriguing new avenues. To utilize this in practice, one could imagine making localized cosmographic reconstructions by partitioning the data into finite patches in redshift and (when data permits) angular space. Each patch should be chosen such that the expansion coefficients can be robustly constrained in it. This division into patches amounts to performing an implicit smoothing whereby angular and radial/redshift scales smaller than the patch size are unresolved. Equivalently, this can be understood as effectively introducing a cutoff in multipole space, ensuring that only a finite number of coefficients need to be constrained. Within a given patch, the primary challenge is then to interpret the physical meaning of the coefficients. One option is to construct suitable combinations of coefficients such as $\mathcal{M}, \mathcal{O}$ and $\mathcal{C}$ defined in \cite{PRL}, which can be directly interpreted as encoding interpretable information. Furthermore, it may be useful to understand if any of the terms in the coefficients are expected to be small, thereby enabling an approximate geometrical or dynamical interpretation of the coefficients themselves. 
\newline\indent
As an example, we will consider each term entering $d_A''$ evaluated along a fiducial light ray for each of our two observers in the LTB1 and LTB2 models, respectively. The comparison of these terms is shown in figure \ref{fig:da_zz_terms}. From this we see that, along these two light rays, two of the terms are several orders of magnitude smaller than the third, at least in a limited redshift range. The dominant term is the one proportional to $-\hat\theta (d\mathcal{H}/d\lambda) /\mathcal{H}$. Thus, the second order expansion coefficient is primarily constraining the term proportional to $-\hat\theta (d\mathcal{H}/d\lambda) /\mathcal{H}$. More realistic assessments of the significance of each term in the expansion coefficients may be assessed through realistic cosmological simulations. 

\section{Summary and conclusion}\label{sec:conclusion}
Motivated by the recent work \cite{arbitraryz1, arbitraryz2} and the documented poor accuracy of general cosmographic expansions at third order centred at $z = 0$, we extend the cosmographic expansion of the redshift-distance relation for a general spacetime to expansions centred at arbitrary redshifts. We demonstrate the use of the expansion by applying it to two examples of LTB models, representing a spherically symmetric  underdensity and overdensity, respectively, surrounded by a $\Lambda$CDM spacetime. Similarly to expansions based at $z = 0$, we overall find that the cosmographic expansions truncated at third order diverge from the exact redshift-distance relation for small redshift intervals when the gradients in density around the centre of expansion are steep. However, the accuracy is greatly improved when the expansion is centred in regions with smoother density contrasts. Since density contrasts are generally modest on the scales probed by e.g. supernovae observations, we thus expect the coefficients of the cosmographic expansion to provide meaningful descriptions of the local expansion and geometry on appropriately smoothed scales \cite{Macpherson:2025qec, convergence2}.
\newline\indent
The purpose of investigating the convergence of cosmographic expansions in LTB models is not solely the hope of demonstrating that low-order cosmography provides a good description of exact redshift-distance relations in inhomogenenous cosmological models. High accuracy for extended redshift ranges at already third order truncation would provide evidence that the coefficients obtained from fitting to observational data can reliably be interpreted in terms of the underlying local dynamics and geometry. However, a lack of convergence is also informative. In particular, it highlights that the physical interpretation of the fitted coefficients is not guaranteed solely by the quality of the fit. In standard cosmology based on FLRW models, Taylor expansions are accurate over an extended redshift range (with $|z|<1$) already at, say, third order. This means that the values of $H_0, q_0, j_0$ etc. obtained from the fitted coefficients actually represent the expansion rate, deceleration and jerk parameter {\em at present time}, and {\em not} their average values over the redshift range of the data. Conversely, in inhomogeneous spacetimes it has been established that the cosmographic expansions tend to have much slower convergence (or smaller radius of convergence), and it is therefore not a priori given that the fitted coefficients may be interpreted as representing genuinely local quantities. Instead, they must be interpreted in terms of (implicit) averages over larger  regions of spacetime/along the lightcone (see e.g. \cite{convergence2, Macpherson:2025qec} for studies of relating the fitted coefficients to spatial averages).
\newline\indent
Cosmographic expansions centred at arbitrary redshift lose the finite multipole structure that make the expansions around $z = 0$ particularly tractable. Nonetheless, the expansions centred at arbitrary redshift provide a flexible framework for constructing local approximations of the redshift-distance relation anywhere along the past light cone. This opens the possibility of localized cosmographic reconstructions and may e.g. increase the precision of model-independent consistency tests of the standard cosmological model. In practice, such analyses would naturally rely on a finite number of redshift and angular patches introducing an effective smoothing scale and limiting the number of free parameters. As part of developing such a formalism, an important direction for future work is to develop a clear physical interpretation of these localized expansion coefficients in terms of the scales they probe, and assess their observational significance, e.g., using realistic simulations.

\acknowledgments
This project was funded by Villum Fonden, grant VIL53032 (PI: SMK).\\
{\bf Author contribution statement:} The presented work is based on results obtained by JBB as part of his master's thesis project conducted under the supervision of SMK who developed the broader research programme and conceptual framework underlying this work, building on the earlier work \cite{PRL} with AH. JBB carried out the numerical and analytical work under the guidance of SMK and, for the analytical results, with independent checks by AH. The analytical results build on initial considerations of AH and SMK made in relation to the work presented in \cite{PRL}. The writing of the manuscript was lead by JBB and SMK, with significant comments and suggestions from AH.\\
{\bf AI declaration:} The numerical work was obtained with code that was written with assistance from large language models (LLMs). The LLMs assisted with minor tasks such as improving figure layout and code efficiency. No LLM suggestions were implemented without scrutinization by JBB. An LLM was used for conducting an extra check of the analytical expression for $d_A'''$. The expression was first derived by JBB and independently verified by AH.
\clearpage
\appendix
\section{FLRW limit of the Taylor expansion of the angular diameter distance}\label{app:FLRWlimit}
In this appendix, we consider the FLRW limit of the expansion of $d_A$ presented in the main text. We first consider $d_A'$. Using that $\mathcal{H}$ in this limit reduces to the Hubble parameter, and that $\hat{\theta}_{\mathrm{FLRW}} = \frac{2}{d_{A, \mathrm{FLRW}}} \frac{\mathrm{d} d_{A, \mathrm{FLRW}}}{\mathrm{d} \lambda} = 2H(1+z)E_0 - \frac{2E_0(1+z)}{d_{A, \mathrm{FLRW}}}$, we find
\begin{align}
    d_{A,\rm FLRW}' =  -\frac{d_{A,\mathrm{FLRW}}}{1+z} + \frac{1}{(1+z)H}.
\end{align}
To identify the FLRW limit of $d_A''$ we note that the shear now vanishes and that
\begin{align}
    R_{\mu\nu}k^\mu k^\nu = 8\pi G(\rho + p)E_o^2(1+z)^2 = 2E_o^2H^2 (1+q)(1+z)^2,
\end{align}
in the FLRW limit, where $q:= -\ddot a/a/H^2$ is the FLRW deceleration parameter. This also means that $H'=H(1+q)/(1+z)$. With these limits at hand, we see that the expression for $d_A''$ in the FLRW limit reduces to
\begin{align}
    d_{A,\mathrm{FLRW}}'' &= \frac{2d_{A,\mathrm{FLRW}}}{(1+z)^2}  - \frac{3+q}{(1+z)^2H}.
\end{align}

Lastly, we consider the FLRW limit of $d_A'''$ where we note that the Weyl curvature vanishes and that
\begin{align}
    H'' = \frac{H}{(1+z)^2}(j-q^2),
\end{align}
and
\begin{align}
    (R_{\mu\nu}k^\mu k^\nu)' = 2E_o^2 H^2 (1+z)(j+5q+4),
\end{align}
where $j:=\dddot a/a/H^3$ is the jerk parameter.\newline\indent
Inserting all the FLRW limits summarized above into the general expression for $d_A'''$ , we see that it, in the FLRW limit, reduces to
\begin{align}
    d_{A, \rm FLRW} ''' = -\frac{6d_{A, \rm FLRW}}{(1+z)^3} + \frac{3q^2 + 7q -j + 11}{(1+z)^3 H}.
\end{align}


\clearpage

\bibliography{manuscript}

@book{god_bog,
    author    = "Ellis, George F. R. and Maartens, Roy and MacCallum, Malcolm A. H.",
    title     = "{Relativistic Cosmology}",
    publisher = "Cambridge University Press",
    address   = "Cambridge",
    year      = "2012",
    isbn      = "9781107603499"
}

@article{tension1,
    author = "Riess, Adam G.",
    title = "{The Expansion of the Universe is Faster than Expected}",
    eprint = "2001.03624",
    archivePrefix = "arXiv",
    primaryClass = "astro-ph.CO",
    doi = "10.1038/s42254-019-0137-0",
    journal = "Nature Rev. Phys.",
    volume = "2",
    number = "1",
    pages = "10--12",
    year = "2019"
}

@article{tension2,
    author = "Lodha, K. and others",
    collaboration = "DESI",
    title = "{Extended dark energy analysis using DESI DR2 BAO measurements}",
    eprint = "2503.14743",
    archivePrefix = "arXiv",
    primaryClass = "astro-ph.CO",
    reportNumber = "FERMILAB-PUB-25-0164-PPD",
    doi = "10.1103/w4c6-1r5j",
    journal = "Phys. Rev. D",
    volume = "112",
    number = "8",
    pages = "083511",
    year = "2025"
}

@article{tension3,
    author = "Wang, Deng and Mota, David",
    title = "{Did DESI DR2 truly reveal dynamical dark energy?}",
    eprint = "2504.15222",
    archivePrefix = "arXiv",
    primaryClass = "astro-ph.CO",
    doi = "10.1140/epjc/s10052-025-15076-y",
    journal = "Eur. Phys. J. C",
    volume = "85",
    number = "11",
    pages = "1356",
    year = "2025"
}

@article{tension4,
    author = "Cort{\^e}s, Marina and Liddle, Andrew R.",
    title = "{On DESI's DR2 exclusion of $\Lambda$CDM}",
    eprint = "2504.15336",
    archivePrefix = "arXiv",
    primaryClass = "astro-ph.CO",
    doi = "10.1093/mnrasl/slaf108",
    journal = "Mon. Not. Roy. Astron. Soc.",
    volume = "544",
    pages = "L121--L125",
    year = "2025"
}

@article{tension5,
    author = "Secrest, Nathan and von Hausegger, Sebastian and Rameez, Mohamed and Mohayaee, Roya and Sarkar, Subir",
    title = "{Colloquium: The cosmic dipole anomaly}",
    eprint = "2505.23526",
    archivePrefix = "arXiv",
    primaryClass = "astro-ph.CO",
    doi = "10.1103/9ygx-z2yq",
    journal = "Rev. Mod. Phys.",
    volume = "97",
    number = "4",
    pages = "041001",
    year = "2025"
}

@article{Kalbouneh:2022tfw,
    author = "Kalbouneh, Basheer and Marinoni, Christian and Bel, Julien",
    title = "{Multipole expansion of the local expansion rate}",
    eprint = "2210.11333",
    archivePrefix = "arXiv",
    primaryClass = "astro-ph.CO",
    doi = "10.1103/PhysRevD.107.023507",
    journal = "Phys. Rev. D",
    volume = "107",
    number = "2",
    pages = "023507",
    year = "2023"
}

@article{Adamek:2024hme,
    author = "Adamek, Julian and Clarkson, Chris and Durrer, Ruth and Heinesen, Asta and Kunz, Martin and Macpherson, Hayley J.",
    title = "{Towards Cosmography of the Local Universe}",
    eprint = "2402.12165",
    archivePrefix = "arXiv",
    primaryClass = "astro-ph.CO",
    doi = "10.33232/001c.118782",
    journal = "Open J. Astrophys.",
    volume = "7",
    pages = "001c.118782",
    year = "2024"
}

@article{pade,
    author = "Wei, Hao and Yan, Xiao-Peng and Zhou, Ya-Nan",
    title = "{Cosmological Applications of Pad{\'e} Approximant}",
    eprint = "1312.1117",
    archivePrefix = "arXiv",
    primaryClass = "astro-ph.CO",
    doi = "10.1088/1475-7516/2014/01/045",
    journal = "JCAP",
    volume = "01",
    pages = "045",
    year = "2014"
}

@article{Etherington:1933,
    author  = "Etherington, I. M. H.",
    title   = "{On the Definition of Distance in General Relativity}",
    journal = "General Relativity and Gravitation",
    volume  = "39",
    pages   = "1055--1067",
    year    = "2007",
    note    = "Reprint of Phil. Mag. (Series 7) 15, 761--773 (1933)",
    doi     = "10.1007/s10714-007-0488-4"
}

@article{arbitraryz1,
    author = "Liu, Yang and Wang, Bao and Yu, Hongwei and Wu, Puxun",
    title = "{Probing cosmic background dynamics with a cosmological-model-independent method}",
    eprint = "2305.19634",
    archivePrefix = "arXiv",
    primaryClass = "astro-ph.CO",
    doi = "10.1093/mnras/stae1808",
    journal = "Mon. Not. Roy. Astron. Soc.",
    volume = "533",
    number = "1",
    pages = "244--253",
    year = "2024"
}

@article{arbitraryz2,
    author = "Fazzari, Elisa and Giar{\`e}, William and Di Valentino, Eleonora",
    title = "{Cosmographic Footprints of Dynamical Dark Energy}",
    eprint = "2509.16196",
    archivePrefix = "arXiv",
    primaryClass = "astro-ph.CO",
    doi = "10.3847/2041-8213/ae2917",
    journal = "Astrophys. J. Lett.",
    volume = "996",
    number = "1",
    pages = "L5",
    year = "2026"
}

@article{convergence1,
    author = "Modan, Asha B. and Koksbang, S. M.",
    title = "{On the convergence of cosmographic expansions in Lema{\^\i}tre{\textendash}Tolman{\textendash}Bondi models}",
    eprint = "2408.07459",
    archivePrefix = "arXiv",
    primaryClass = "gr-qc",
    doi = "10.1088/1361-6382/ad8abc",
    journal = "Class. Quant. Grav.",
    volume = "41",
    number = "23",
    pages = "235018",
    year = "2024"
}

@article{convergence2,
    author = "Koksbang, S. M.",
    title = "{Testing inhomogeneous cosmography in our cosmic neighborhood using CosmicFlows-4}",
    eprint = "2412.12637",
    archivePrefix = "arXiv",
    primaryClass = "astro-ph.CO",
    doi = "10.1103/6z7w-47rc",
    journal = "Phys. Rev. D",
    volume = "111",
    number = "12",
    pages = "123516",
    year = "2025"
}

@article{convergence3,
    author = "Hills, Morag and Heinesen, Asta",
    title = "{Cosmography with {\ensuremath{\Lambda}}-Szekeres models}",
    eprint = "2601.16844",
    archivePrefix = "arXiv",
    primaryClass = "astro-ph.CO",
    doi = "10.1088/1475-7516/2026/06/016",
    journal = "JCAP",
    volume = "06",
    pages = "016",
    year = "2026"
}

@article{weaklensing1,
    author = "Hu, Wayne and Keeton, Charles R.",
    title = "{Three-dimensional mapping of dark matter}",
    eprint = "astro-ph/0205412",
    archivePrefix = "arXiv",
    doi = "10.1103/PhysRevD.66.063506",
    journal = "Phys. Rev. D",
    volume = "66",
    pages = "063506",
    year = "2002"
}

@article{galaxy,
    author = "Quevedo, B. Camacho and others",
    collaboration = "Euclid",
    title = "{Euclid preparation. Galaxy power spectrum modelling in redshift space}",
    eprint = "2601.20826",
    archivePrefix = "arXiv",
    primaryClass = "astro-ph.CO",
    month = "1",
    year = "2026"
}

@article{velocity,
    author = "Turner, Ryan J.",
    title = "{Cosmology with Peculiar Velocity Surveys}",
    eprint = "2411.19484",
    archivePrefix = "arXiv",
    primaryClass = "astro-ph.CO",
    month = "11",
    year = "2024"
}

@article{CosmicFlows,
    author = "Courtois, H. M. and Dupuy, A. and Guinet, D. and Baulieu, G. and Ruppin, F. and Brenas, P.",
    title = "{Gravity in the Local Universe : density and velocity fields using CosmicFlows-4}",
    eprint = "2211.16390",
    archivePrefix = "arXiv",
    primaryClass = "astro-ph.CO",
    doi = "10.1051/0004-6361/202245331",
    journal = "Astron. Astrophys.",
    volume = "670",
    pages = "L15",
    year = "2023"
}

@article{Riess:2016jrr,
    author = "Riess, Adam G. and others",
    title = "{A 2.4{\%} Determination of the Local Value of the Hubble Constant}",
    eprint = "1604.01424",
    archivePrefix = "arXiv",
    primaryClass = "astro-ph.CO",
    doi = "10.3847/0004-637X/826/1/56",
    journal = "Astrophys. J.",
    volume = "826",
    number = "1",
    pages = "56",
    year = "2016"
}

@article{Riess:2021jrx,
    author = "Riess, Adam G. and others",
    title = "{A Comprehensive Measurement of the Local Value of the Hubble Constant with 1 km s$^{-1}$ Mpc$^{-1}$ Uncertainty from the Hubble Space Telescope and the SH0ES Team}",
    eprint = "2112.04510",
    archivePrefix = "arXiv",
    primaryClass = "astro-ph.CO",
    doi = "10.3847/2041-8213/ac5c5b",
    journal = "Astrophys. J. Lett.",
    volume = "934",
    number = "1",
    pages = "L7",
    year = "2022"
}

@article{Dhawan:2022yws,
    author = "Dhawan, Suhail and others",
    title = "{A Uniform Type Ia Supernova Distance Ladder with the Zwicky Transient Facility: Absolute Calibration Based on the Tip of the Red Giant Branch Method}",
    eprint = "2203.04241",
    archivePrefix = "arXiv",
    primaryClass = "astro-ph.CO",
    doi = "10.3847/1538-4357/ac7ceb",
    journal = "Astrophys. J.",
    volume = "934",
    number = "2",
    pages = "185",
    year = "2022"
}

@article{Galbany:2022zir,
    author = "Galbany, Llu{\'\i}s and others",
    title = "{An updated measurement of the Hubble constant from near-infrared observations of Type Ia supernovae}",
    eprint = "2209.02546",
    archivePrefix = "arXiv",
    primaryClass = "astro-ph.CO",
    doi = "10.1051/0004-6361/202244893",
    journal = "Astron. Astrophys.",
    volume = "679",
    pages = "A95",
    year = "2023"
}

@article{Uddin:2023iob,
    author = "Uddin, Syed A. and others",
    title = "{Carnegie Supernova Project I and II: Measurements of H $_{0}$ Using Cepheid, Tip of the Red Giant Branch, and Surface Brightness Fluctuation Distance Calibration to Type Ia Supernovae*}",
    eprint = "2308.01875",
    archivePrefix = "arXiv",
    primaryClass = "astro-ph.CO",
    doi = "10.3847/1538-4357/ad3e63",
    journal = "Astrophys. J.",
    volume = "970",
    number = "1",
    pages = "72",
    year = "2024"
}

@article{Riess:2022oxy,
    author = "Riess, Adam G. and Breuval, Louise",
    title = "{The Local Value of H0}",
    eprint = "2308.10954",
    archivePrefix = "arXiv",
    primaryClass = "astro-ph.CO",
    doi = "10.1017/S1743921323003034",
    journal = "IAU Symp.",
    volume = "376",
    pages = "15--29",
    year = "2022"
}

@article{DES:2024ywx,
    author = "Camilleri, R. and others",
    collaboration = "DES",
    title = "{The Dark Energy Survey Supernova Program: an updated measurement of the Hubble constant using the inverse distance ladder}",
    eprint = "2406.05049",
    archivePrefix = "arXiv",
    primaryClass = "astro-ph.CO",
    reportNumber = "DES-2024-835, FERMILAB-PUB-24-0290-PPD",
    doi = "10.1093/mnras/staf122",
    journal = "Mon. Not. Roy. Astron. Soc.",
    volume = "537",
    number = "2",
    pages = "1818--1825",
    year = "2025"
}

@article{Clarkson:2011uk,
    author = "Clarkson, Chris and Umeh, Obinna",
    title = "{Is backreaction really small within concordance cosmology?}",
    eprint = "1105.1886",
    archivePrefix = "arXiv",
    primaryClass = "astro-ph.CO",
    doi = "10.1088/0264-9381/28/16/164010",
    journal = "Class. Quant. Grav.",
    volume = "28",
    pages = "164010",
    year = "2011"
}

@phdthesis{umeh2013,
    author = "Umeh, Obinna",
    title = "{The influence of structure formation on the evolution of the universe}",
    school = "University of Cape Town",
    year = "2013",
    type = "PhD thesis",
    address = "Cape Town, South Africa",
    url = "https://open.uct.ac.za/handle/11427/32662"
}

@article{Heinesen:2020bej,
    author = "Heinesen, Asta",
    title = "{Multipole decomposition of the general luminosity distance 'Hubble law' -- a new framework for observational cosmology}",
    eprint = "2010.06534",
    archivePrefix = "arXiv",
    primaryClass = "astro-ph.CO",
    doi = "10.1088/1475-7516/2021/05/008",
    journal = "JCAP",
    volume = "05",
    pages = "008",
    year = "2021"
}

@article{Maartens:2023tib,
    author = "Maartens, Roy and Santiago, Jessica and Clarkson, Chris and Kalbouneh, Basheer and Marinoni, Christian",
    title = "{Covariant cosmography: the observer-dependence of the Hubble parameter}",
    eprint = "2312.09875",
    archivePrefix = "arXiv",
    primaryClass = "astro-ph.CO",
    doi = "10.1088/1475-7516/2024/09/070",
    journal = "JCAP",
    volume = "09",
    pages = "070",
    year = "2024"
}

@article{Kalbouneh:2024szq,
    author = "Kalbouneh, Basheer and Marinoni, Christian and Maartens, Roy",
    title = "{Cosmography of the local Universe by multipole analysis of the expansion rate fluctuation field}",
    eprint = "2401.12291",
    archivePrefix = "arXiv",
    primaryClass = "astro-ph.CO",
    doi = "10.1088/1475-7516/2024/09/069",
    journal = "JCAP",
    volume = "09",
    pages = "069",
    year = "2024"
}

@article{Dhawan:2022lze,
    author = "Dhawan, Suhail and Borderies, Antonin and Macpherson, Hayley J. and Heinesen, Asta",
    title = "{The quadrupole in the local Hubble parameter: first constraints using Type Ia supernova data and forecasts for future surveys}",
    eprint = "2205.12692",
    archivePrefix = "arXiv",
    primaryClass = "astro-ph.CO",
    doi = "10.1093/mnras/stac3812",
    journal = "Mon. Not. Roy. Astron. Soc.",
    volume = "519",
    number = "4",
    pages = "4841--4855",
    year = "2023"
}

@article{Cowell:2022ehf,
    author = "Cowell, Jessica A. and Dhawan, Suhail and Macpherson, Hayley J.",
    title = "{Potential signature of a quadrupolar hubble expansion in Pantheon+supernovae}",
    eprint = "2212.13569",
    archivePrefix = "arXiv",
    primaryClass = "astro-ph.CO",
    doi = "10.1093/mnras/stad2788",
    journal = "Mon. Not. Roy. Astron. Soc.",
    volume = "526",
    number = "1",
    pages = "1482--1494",
    year = "2023"
}

@article{Macpherson:2025qec,
    author = "Macpherson, Hayley J. and Heinesen, Asta",
    title = "{A theoretical prediction for the dipole in nearby distances using cosmography}",
    eprint = "2507.01095",
    archivePrefix = "arXiv",
    primaryClass = "astro-ph.CO",
    doi = "10.33232/001c.150319",
    month = "7",
    year = "2025"
}

@article{Lemaitre1933,
    author = "Lema{\^i}tre, Georges",
    title = "{L'Univers en expansion}",
    journal = "Annales de la Soci\'et\'e Scientifique de Bruxelles A",
    volume = "53",
    pages = "51",
    year = "1933"
}

@article{Lemaitre1997,
    author = "Lema{\^i}tre, Georges",
    title = "{The Expanding Universe}",
    journal = "General Relativity and Gravitation",
    volume = "29",
    pages = "637",
    year = "1997",
    note = "English translation of Lema{\^i}tre (1933)"
}

@article{Tolman1934,
    author = "Tolman, Richard C.",
    title = "{Effect of Inhomogeneity on Cosmological Models}",
    journal = "Proceedings of the National Academy of Sciences",
    volume = "20",
    pages = "169--176",
    year = "1934"
}

@article{Bondi1947,
    author = "Bondi, Hermann",
    title = "{Spherically Symmetrical Models in General Relativity}",
    journal = "Monthly Notices of the Royal Astronomical Society",
    volume = "107",
    pages = "410--425",
    year = "1947"
}

@misc{PRL,
  author = {Koksbang, S. M. and Heinesen, A.},
  title  = "{Diagnostic Consistency Tests of the Concordance Cosmology}",
  year   = {2026},
  archivePrefix= "arXiv"
}

@book{Bolejko2010,
year = {2010},
author = {Bolejko, Krzysztof},
address = {Cambridge},
booktitle = {Structures in the universe by exact methods : formation, evolution, interactions},
isbn = {1-107-20936-6},
language = {eng},
publisher = {Cambridge University Press},
series = {Cambridge monographs on mathematical physics},
title = {Structures in the universe by exact methods : formation, evolution, interactions },
}

@article{VanAcoleyen:2008cy,
    author = "Van Acoleyen, Karel",
    title = "{LTB solutions in Newtonian gauge: From Strong to weak fields}",
    eprint = "0808.3554",
    archivePrefix = "arXiv",
    primaryClass = "gr-qc",
    doi = "10.1088/1475-7516/2008/10/028",
    journal = "JCAP",
    volume = "10",
    pages = "028",
    year = "2008"
}

@article{Koksbang:2021zyi,
    author = "Koksbang, S. M.",
    title = "{Understanding the Dyer-Roeder approximation as a consequence of local cancellations of projected shear and expansion rate fluctuations}",
    eprint = "2106.12913",
    archivePrefix = "arXiv",
    primaryClass = "astro-ph.CO",
    doi = "10.1103/PhysRevD.104.043505",
    journal = "Phys. Rev. D",
    volume = "104",
    number = "4",
    pages = "043505",
    year = "2021"
}

@article{healpix,
    author = "G{\'o}rski, K. M. and Hivon, E. and Banday, A. J. and Wandelt, B. D. and Hansen, F. K. and Reinecke, M. and Bartelmann, M.",
    title = "{HEALPix - A Framework for high resolution discretization, and fast analysis of data distributed on the sphere}",
    eprint = "astro-ph/0409513",
    archivePrefix = "arXiv",
    doi = "10.1086/427976",
    journal = "Astrophys. J.",
    volume = "622",
    pages = "759--771",
    year = "2005"
}

\end{document}